# Design of ultra-high gain optical micro-amplifiers via smart non-linear wave mixing

Özüm Emre Aşırım [1,*] and Alim Yolalmaz [2]

1. Department of Electrical and Electronics Engineering, Middle East Technical University, 06800 Ankara, Turkey
2. Micro and Nanotechnology, Middle East Technical University, 06800 Ankara, Turkey

* Correspondence: e176154@metu.edu.tr

**Abstract**

Optical amplification of the input wave by mixing the pump wave within a nonlinear interaction medium offers high gain for a variety of applications. In real life studies, the interaction mediums which allow the optical amplification of the input wave have many resonance frequencies. However, the computational expense for tuning the pump frequency to yield the optical amplification of the input wave increases with the number of resonance frequencies within the interaction mediums. Here, we present a Fletcher-Reeves based algorithm for parametric amplification in micro-resonators having multiple resonance frequencies. Using our novel mathematical formulations, we obtained a gain of $4.7 \times 10^7$ for the input wave at 640 THz and a gain of $1.5 \times 10^8$ for the input wave at 100 THz within the micro-resonators. Moreover, the performance of our algorithm is verified by the well know mathematical expression, and we achieved more than 99% accuracy in computation of optical amplification. To our knowledge, this is the first study where Fletcher-Reeves algorithm is used for the parametric amplification. Our methodology can be accompanied to design optical parametric amplifiers for applications of high-speed optical communications, photonic circuits, and ultrafast lasers.

---

**Introduction**

Optical parametric amplification via nonlinear wave mixing has been extensively studied[1–7] and can be used for a variety of applications such as photonic integrated circuits[8], high-speed optical comunications[9], piezoelectric MEMS device[10], and development of powerful ultrafast lasers[11]. Optical parametric amplification is achieved via mixing a non-linearity inducing high-intensity pump wave with an input wave of low intensity, during which the low-intensity input wave gets amplified by absorbing energy from the high-intensity pump wave[12,13]. The rate and amount of the energy that can be transferred depending on the non-linearity of the interaction medium[14]; thus, the gain of the parametric amplification depends heavily on the length of the interaction medium. The required interaction medium length for high optical gain is usually on the order of centimeters.

Novel ultra-wideband optical antennas, super-efficient harmonic generators[15], and small-scale optical ablation devices need the optical gain via the non-linear wave mixing process in the micro and nanoscale which is a crucial scientific problem. For achieving significant optical gain in the millimeter or lower scale, specially designed materials showing nonlinear optical properties are required[16–18]. Therefore, parametric amplifiers are currently not feasible to be employed in the micro and nanometer scale for high-gain optical amplification with ordinary materials. Optical parametric amplification is mostly studied experimentally[10,19–22]. There are a few computational studies where optimization of pump frequency is investigated[5,6,14,23], but the interaction mediums have a single resonance frequency. However, in real applications, the interaction mediums of nonlinear wave mixers have multiple resonance frequencies which causes high computation cost.

In this study, we design Fabry-Perot micro-resonators to amplify a wave at 640 THz and a wave at 100 THz separately. The micro-resonators we use here consist of multiple resonance frequencies which mimic real materials property. Using our novel mathematical formulations for the nonlinear wave mixing of the pump wave with the input wave and the Fletcher-Reeves algorithm we search frequency of the pump wave to obtain high gain at the desired frequencies of waves. Our results show that the Fletcher-Reeves algorithm and our formulations enable us to generate high gain optical amplification within the micro-resonators with multiple resonance frequencies. Moreover, the gain factors we get here show similarity with the results of well-known experimental formulas.

**Methods**

The configuration of a Fabry-Perot micro-resonator used here is illustrated in Fig. 1. The micro-resonator has an optical isolator at the input port and a band-pass filter at the output port. We consider an interaction medium with 10 μm length having 3 resonance frequencies associated with 3 corresponding polarization damping rates. The pump waves consisting of two ultra-short waves and the low-intensity input wave are excited from the optical isolator wall.

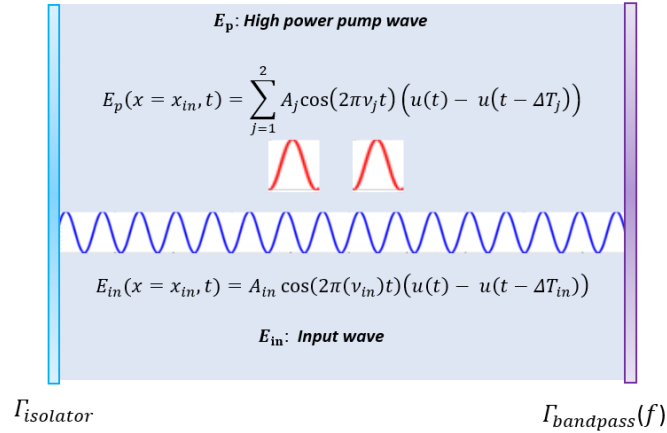

**Fig. 1 A scheme of Fabry-Perot micro-resonator to be used in the numerical simulations.** The micro-resonator is excited by two ultra-short pump waves indicated by red pulses and one low-intense input wave indicated by blue pulse trains. The optical isolator is at the input port (left); the band-pass filter is at the output port (right).

The propagation of the pump waves is described by Eqs. (1-2).[5,23] Eq. (1) represents the electric field wave equation of the pump wave $E_p(\nu)$ at pump frequency $\nu$. The variables in Eq. (1) are $\mu_0$: magnetic permeability, $\varepsilon_\infty$: background permittivity, t: time, σ: medium electrical conductivity, and $P_p(\nu)$: total polarization density generated by the pump wave at pump frequency $\nu$. Eq. (2) represents the associated polarization density components induced by the pump wave $P_{p,i}(\nu_i)$ for ith resonance frequency $\nu_i$. $\gamma_i$: polarization decay rate of the ith resonance frequency, $\omega_i$: ith angular resonance frequency, e: electron charge, d: atomic diameter of 0.3 nm, m: mass of an electron, and $Q_i$: the number of electrons of the ith resonance frequency per unit volume. Note that each resonance frequency $\nu_i$ is associated with a different polarization decay rate, electron density, and polarization density component. Therefore, the total polarization density in Eq. (1), is the sum of all polarization density components indicated in Eq. (2).

$$\nabla^2\big(E_p(\nu)\big) - \mu_0\varepsilon_\infty \frac{\partial^2\big(E_p(\nu)\big)}{\partial t^2} = \mu_0\sigma \frac{\partial\big(E_p(\nu)\big)}{\partial t} + \mu_0 \frac{d^2 P_p(\nu)}{\partial t^2}, \qquad (1)$$

$$\frac{d^2 P_{p,i}(\nu_i)}{dt^2} + \gamma_i \frac{dP_{p,i}(\nu_i)}{dt} + \omega_i^2 P_{p,i}(\nu_i) - \frac{\omega_i^2 P_{p,i}(\nu_i)^2}{Q_i ed} + \frac{\omega_i^2 P_{p,i}(\nu_i)^3}{Q_i^2 e^2 d^2} = \frac{Q_i e^2 E_p(\nu)}{m}, \qquad (2)$$

Resonance probability at the ith resonance of the medium $\xi_i$ expressed by Eq. (3) plays roles in nonlinear wave mixing, and Quantum mechanics dictates that the sum of resonance probabilities is equal to 1[24,25].

Therefore, for a given electromagnetic wave in the medium of propagation, the relation between the resonance probability $\xi_i$ and the associated polarization density component $P_i$ at ith resonance frequency is expressed by Eq. (3), and a sum of all polarization density components is equal to total polarization density $P_p(v)$ in Eq. (1). $p_i$ in Eq. (3) is the induced electric dipole moment contributed by the ith resonance frequency of the medium, and Q is the total number of electrons per unit volume as $3.5 \times 10^{28}$ m$^{-3}$.

$$P_i = Q \sum_{i=1}^{k} \xi_i * p_i, \tag{3}$$

When both the low-intensity input wave $E_{in}$ and the high-intensity pump wave $E_p$ are present in the same interaction medium, the electrical wave field and the associated polarization density can be expressed by using Eqs. (1-2) with summed electric field ($E_p + E_{in}$) and summed polarization density ($P_p + P_{in}$). By subtracting equations derived for the sum of pump wave $E_p$ and the low-intensity wave $E_{in}$ from equations for only the pump wave $E_p$, we get the equations that describe the propagation of the low-intensity input wave under the presence of the high-intensity pump wave as Eqs. (4-5). As we can see in Eq. 5, induced polarization density $P_p$ by the pump wave is coupled to the input wave polarization density $P_{in}$. Therefore, the polarization density that is induced by the pump wave acts as a coupling coefficient arising from the nonlinearity.

$$\nabla^2(E_{in}) - \mu_0 \varepsilon_\infty \frac{\partial^2(E_{in})}{\partial t^2} = \mu_0 \sigma \frac{\partial(E_{in})}{\partial t} + \mu_0 \frac{d^2(P_{in}(v))}{\partial t^2}, \tag{4}$$

$$\frac{d^2 P_{in,i}(v)}{dt^2} + \gamma_i \frac{dP_{in,i}(v)}{dt} + \omega_i^2(P_{in,i}(v)) + \frac{\omega_i^2(P_{in,i}(v)^2 + 2P_{p,i}(v)P_{in,i}(v))}{Q_i ed} \\ - \frac{\omega_i^2(P_{in,i}(v)^3 + 3P_{p,i}(v)P_{in,i}(v)^2 + 3P_{p,i}(v)^2 P_{in,i}(v))}{Q_i^2 e^2 d^2} = \frac{Q_i e^2 (E_{in}(v))}{m}, \tag{5}$$

Considering the design of the Fabry-Perot micro-resonator, the left wall of the resonator blocks any transmission of the pulse from its right side. In that sense, once a pulse enters the cavity through the isolator side, it does not go out of the cavity from the isolator side. The right cavity wall is an optical bandpass filter which filters wave within the micro-resonator. The frequency-dependent magnitude response of the optical bandpass filter $|\Gamma(v)|$ located at the output wall of the micro-resonator is expressed with Eq. 6. $v'$ is the frequency of wave within the micro-resonator, $v$ is the center frequency of the desired wave, and $\Delta v$ is a half-bandwidth value of the desired output frequency centered around the frequency of the amplified input wave.

$$|\Gamma(v)| = 1 - e^{-(\frac{(v-v')}{\sqrt{2\Delta v^2}})^2}, \tag{6}$$

We define the electric field of the pump wave at the excitation point to be a combination of two intense ultrashort pulses with Eq. 7 where $A_j$: jth wave amplitude, $v_j$: frequency of the jth wave, $u(t)$: unit step function, $\Delta T_j$: pulse duration of the jth wave. The electric field of input wave to be amplified at the excitation point is expressed with Eq. 8. The parameters in Eq. 8 are $A_{in}$: amplitude of the input wave (1 V/m), $v_{in}$: the input wave frequency, $\Delta T_{in}$: pulse duration of the input wave.

$$E_p(x,t) = \sum_{j=1}^{2} A_j \cos(2\pi v_j t)[u(t) - u(t - \Delta T_j)], \tag{7}$$

$$E_{in}(x,t) = A_{in} \cos(2\pi v_{in} t)[u(t) - u(t - \Delta T_{in})], \tag{8}$$

Here, we maximize the absolute value of the peak amplitude of the input wave at $v_{in}$ 640 THz (blue light) with a half-width at half maximum of 10 THz. Three resonance frequencies of the medium are as follows: $v_1$: 380 THz, $v_2$: 540 THz, and $v_3$: 750 THz. Polarization damping rates of the medium at three resonance frequencies are 0.5 THz, 1 THz, and 3 THz, respectively. Resonance strengths are 1/3, 4/9, and 2/9 for three resonance frequencies, respectively. The relative permittivity of the interaction medium $\varepsilon_r$ is 10, and permeability of the interaction medium $\mu_r$ is 1. Parameters of the two pump sources are as follows: $A_1$: $1\times10^8$ V/m, $A_2$: $8\times10^7$ V/m, $\Delta T_1$: 5 ps, and $\Delta T_2$: 8 ps. The passband filter is centered at 640 THz (blue light filter) and has a Gaussian frequency selectivity curve with a half-bandwidth value of 10 THz.

After defining the equations of electric field and polarization density with Eqs. (1,2,4,5), parameters of the micro-resonator, input wave and the pump waves, we discretized Eqs. (1,2,4,5) to compute the electric field through the micro-resonator. These equations are solved using the finite difference time domain (FDTD) method at every update of the non-linear programming process. Since the problem is non-linear, the stability condition is stricter; therefore, it is best to choose the temporal and spatial sampling periods ($\Delta t, \Delta x$) to be as small as possible for more accurate solutions. The simulation is performed for 15 ps. The initial conditions of the electric field, the polarization density, derivative of the electric field with respect to position, and derivative of the polarization density with respect to position is zero.

The cost function F as indicated in Eq. 9 is the maxima of the absolute value of the input wave spectral density around the desired frequency 640 THz in given half-bandwidth $\Delta v$ of 10 THz. Since our problem is a constrained maximization problem, the frequencies of the two pump waves $v_1$ and $v_2$ can only be tuned in a certain frequency range between 100 THz and 400 THz. We update the cost function by adding penalty coefficients $\delta_i$ to decrease it. For each pump wave frequency, there are two penalty coefficients each of which is decided by the upper-frequency range limit and the lower frequency range limit. If the frequency of first pump wave is higher than the lower frequency range limit, first penalty coefficient of the first pump wave is zero; otherwise, first penalty coefficient of the first pump wave is $(1-\frac{1}{\zeta})/(\Delta\Omega)^2$. If the frequency of the first pump wave is lower than the upper-frequency range limit, second penalty coefficient of first pump wave is zero; otherwise, second penalty coefficient of first pump wave is $(1-\frac{1}{\zeta})/(\Delta\Omega)^2$ where reduction factor $\zeta$ (1.5) and a factor of deviation from the max/min allowable frequency $\Delta\Omega$ (10 THz). In case of violation of the constraints, we enforce the satisfaction of the constraints for a maximization based non-linear programming problem. Penalty coefficients of the second pump wave are decided with a similar manner.

$$F = \left| \int_{v_{target}-\Delta v}^{v_{target}+\Delta v} \left\{ \int_0^{\Delta T} \{E_{in}(x=x',t)e^{-i(2\pi\Omega)t}\}dt \right\} e^{i(2\pi\Omega)t} d\Omega \right|_{max} - \delta_1(v_1-v_{max})^2 \\ - \delta_2(v_1-v_{min})^2 - \delta_3(v_2-v_{max})^2 - \delta_4(v_2-v_{min})^2 \tag{9}$$

To maximize the gain and amplitude of the input wave at 640 THz with a half-width at half maximum of 10 THz, we tune the frequencies of the ultrashort intense excitation pulses $v_1$ and $v_2$ via non-linear programming using the Fletcher-Reeves algorithm. The Fletcher Reeves algorithm is a very convenient algorithm for solving non-linear conjugate gradient programming problems[26–29]. Hence, we used the Fletcher-Reeves algorithm as the optimization algorithm for input wave gain-factor maximization.

The Fletcher-Reeves algorithm works as follows: considering the initial ultrashort pulse frequencies $\boldsymbol{v_0}$ (the bold symbol indicates vector; 0 means the starting value), the cost function $F(\boldsymbol{v_0})$, and its gradient $\nabla F(\boldsymbol{v_0})$ are evaluated. If the gradient is not lower than $10^{-5}$ (we chose this value considering simulation time), we compute the step size $\alpha_k$ using the backtracking line search. The backtracking line search starts with selecting initial values of c and $\rho$ between 0 and 1, and the initial value of $\alpha_0$ is selected as a number greater than 0. Then, the initial search direction $\boldsymbol{p_0}$ is chosen as the identity matrix. Next, we update $\alpha_k$ as $\alpha_k\rho$ if $F(\boldsymbol{v_k}+\alpha_k\boldsymbol{p_k}) \leq F(\boldsymbol{v_k})+c\alpha_k\nabla F_k^T\boldsymbol{p_k}$ is satisfied, and we decrease $\alpha_k$ until sufficient decrease condition is reached. Later, ultrashort pulse frequencies at (k+1)th iteration $\boldsymbol{v_{k+1}}$ is computed with $\boldsymbol{v_{k+1}} = \boldsymbol{v_k} + \alpha_k \boldsymbol{p_k}$. Then the gradient of the cost function at (k+1)th $\nabla F_{k+1}$ is calculated by using $\beta_{k+1}$ (see Eq. 10). Next, the search direction at (k+1)th iteration $\boldsymbol{p_{k+1}}$ is updated with $-\nabla F_{k+1} + \beta_{k+1} \boldsymbol{p_k}$. Then the algorithm updates the cost function as the gradient of the cost function $\nabla F(\boldsymbol{v_k})$ at each iteration, and calculation of the cost function gradient continues until the gradient becomes lower than $10^{-5}$.

$$\beta_{k+1} = \frac{\nabla F_{k+1}{}^T \nabla F_{k+1}}{\nabla F_k{}^T \nabla F_k} \tag{10}$$

**Results**

As illustrated in Table 1, the gain factor of the input wave at 640 THz is proportional to the intracavity electric energy density $W_{e,p}$ and the polarization density $P_p$ induced by the pump wave. Therefore, we conclude that the input wave gain-factor maximization problem is equivalent to the concurrent maximization of the intracavity electric energy density and the corresponding polarization density, created by the pump wave. This is expected, as the accumulated polarization density signifies the strength of non-linear coupling, hence, maximizing the electric energy alone is not sufficient to amplify the input wave. As a result, a high-gain amplification of $4.7 \times 10^7$ is achieved at the 42$^{nd}$ update.

**Table 1 Variation of two pump frequencies with the micro-resonator parameters through the algorithm iterations for the generation of a wave at 640 THz.**

| $v_1$ (THz) | $v_2$ (THz) | $Gain_{max}$ | $W_{e,p}$ (J/m$^3$) | $P_p$ (C/m$^2$) | k (iteration #) |
|---|---|---|---|---|---|
| 270 | 260.0 | 1.23 | $1.5 \times 10^7$ | 0.12 | 1 |
| 241.6 | 252.8 | 10.29 | $2.9 \times 10^7$ | 0.13 | 5 |
| 218.9 | 278.2 | 41.84 | $4.4 \times 10^7$ | 0.13 | 9 |
| 238.5 | 288.1 | 117.51 | $6.9 \times 10^7$ | 0.14 | 13 |
| 204.5 | 302.2 | 3255.34 | $9.2 \times 10^7$ | 0.16 | 17 |
| 198.4 | 311.6 | 5287.14 | $1.5 \times 10^8$ | 0.16 | 21 |
| 172.4 | 288.5 | $1.46 \times 10^4$ | $2.8 \times 10^8$ | 0.18 | 25 |
| 187.0 | 323.1 | $2.16 \times 10^5$ | $8.8 \times 10^8$ | 0.20 | 29 |
| 156.6 | 349.7 | $8.73 \times 10^5$ | $9.7 \times 10^8$ | 0.22 | 33 |
| 164.8 | 352.4 | $5.68 \times 10^6$ | $1.6 \times 10^9$ | 0.24 | 37 |
| 162.3 | 351.3 | $3.30 \times 10^7$ | $3.0 \times 10^9$ | 0.26 | 40 |
| 161.7 | 350.9 | $4.70 \times 10^7$ | $3.1 \times 10^9$ | 0.27 | 42 |

In Fig. 2a input wave amplitude within the micro-resonator though optimization time of 15 ps at 42$^{nd}$ iteration is presented. As seen here we obtained around 50 MV/m amplitude at 640 THz. Intensity spectral density after the bandpass filter shows that a wave at 640 THz is generated (Fig. 2b). As a result of this study, we conclude that our formulations and the Fletcher-Reeves algorithm enable us to optimize the pump wave frequencies to amplify the input wave to the gain of $4.7 \times 10^7$.

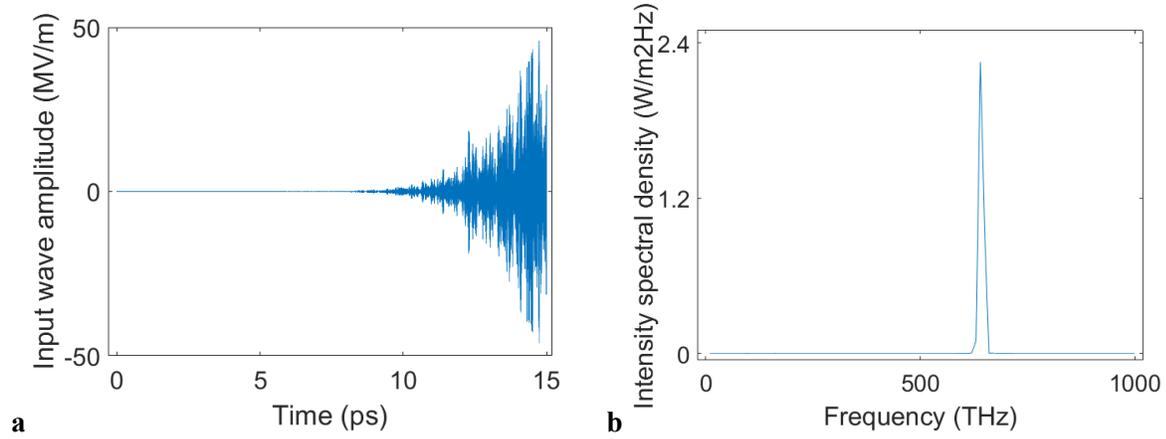

**Fig. 2 Results of wave-amplification for the generation of a wave at 640 THz. a** Amplitude variation of the input wave through optimization time of 15 ps at the bandpass filter output; **b** Spectrum of the generated wave at the bandpass filter output.

After amplifying the wave at 640 THz, we perform a similar study in order to intensify the wave at 100 THz with 20 THz FHWM using the same methodology. In this study, we maximize the amplitude of the wave at 100 THz with a single pump frequency within the micro-resonator having two resonance frequencies. Our interaction medium has two resonance frequencies as $\nu_1$: 1100 THz and $\nu_2$: 1500 THz. Polarization damping rates of the medium at two resonance frequencies are 2 THz and 5 THz, respectively. Resonance strength at each resonance frequency is 0.5. The relative permittivity of the interaction medium $\varepsilon_r$ and permeability of the interaction medium $\mu_r$ is 12 and 1, respectively. The amplitude of the input wave $A_{in}$ is 1 V/m, and the amplitude of the pump wave is $A_p$: 1.8x10$^8$ V/m. The simulation is performed during 20 ps. In this study, the pump frequency is restricted between 100 THz and 400 THz. The half-bandwidth value of the optical bandpass filter expressed in Eq. 6 is 5 THz. Then the cost function is expressed after modifying Eq. 9 to optimize and find a single pump frequency considering only two penalty coefficients.

Based on these optimization parameters, the cavity parameters and the pump frequency are summarized in Table 2. From Table 2, again we can see that the gain factor of the input wave at the desired frequency of 100 THz, is proportional to the intracavity electric energy density $W_{e,p}$ and the polarization density induced by the pump wave $P_p$. As stated early, we deduce that the input wave gain-factor maximization problem is equivalent to the concurrent maximization of the intracavity electric energy density and the corresponding polarization density, created by the pump wave. The highest gain factor for the amplification of the wave at 100 THz is achieved at the 28$^{th}$ update as 1.5x10$^8$.

**Table 2 Variation of the cavity parameters and pump wave frequency through the algorithm iterations for the generation of a wave at 100 THz.**

| $\nu_p$ (THz) | $Gain_{max}$ | $W_{e,p}$ (J/m$^3$) | $P_{pump}$ (C/m$^2$) | k (iteration #) |
|---|---|---|---|---|
| 225.0 | 13.2 | $2.4 \times 10^7$ | 0.14 | 1 |
| 257.0 | 17.7 | $2.5 \times 10^7$ | 0.16 | 4 |
| 260.1 | 83.4 | $2.7 \times 10^7$ | 0.17 | 7 |
| 262.5 | 196.0 | $4.1 \times 10^7$ | 0.19 | 10 |
| 264.6 | 1271.6 | $4.9 \times 10^7$ | 0.20 | 12 |
| 259.5 | 6350.8 | $6.6 \times 10^7$ | 0.22 | 14 |
| 227.3 | 2.6× 10$^4$ | $9.3 \times 10^7$ | 0.23 | 16 |
| 83.3 | 5.3× 10$^4$ | $1.1 \times 10^8$ | 0.25 | 18 |
| 143.3 | 3.07× 10$^5$ | $2.8 \times 10^8$ | 0.26 | 20 |
| 158.0 | $1.2\times 10^6$ | $3.7\times 10^8$ | 0.28 | 22 |
| 166.9 | $8.4\times 10^6$ | $1.0\times 10^9$ | 0.29 | 24 |
| 152.9 | $4.1\times 10^7$ | $2.2\times 10^9$ | 0.30 | 26 |
| 157.4 | $1.5\times 10^8$ | $3.2\times 10^9$ | 0.31 | 28 |

Fig. 3a displays the amplification of the input wave around 100 THz (quasi-monochromatic with a bandwidth of 20 THz) with the optimal frequency of the ultrashort pulse as 157.4 THz that leads to the optimal pump wave excitation. A very high gain-factor of $1.5\times 10^8$ at the filter output is achieved after 20 ps. The amplitude of the input wave increases significantly after 10 ps due to the exponential rate of amplification, in compliance with the experimental results[24,25]. In Fig. 3b we present intensity spectral density. The inset shows the input wave at around 100 THz is intensified.

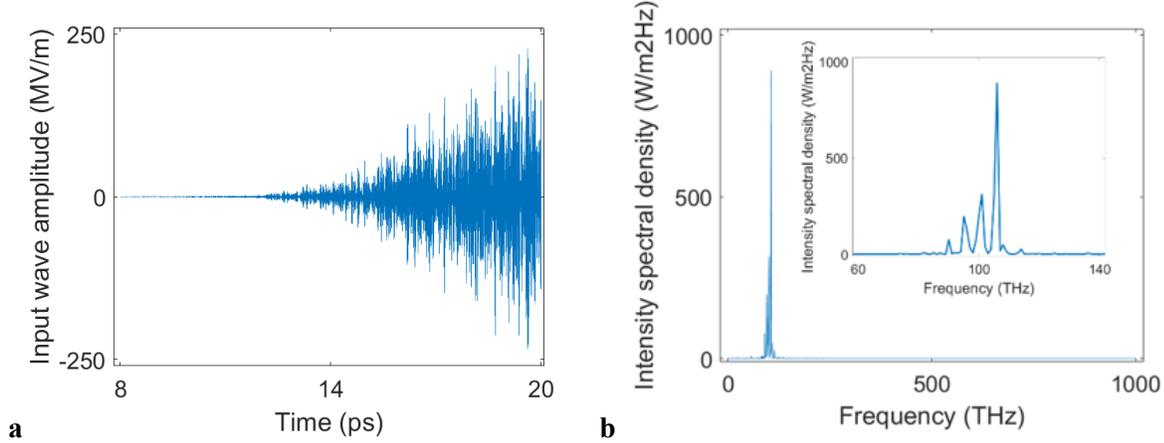

**Fig. 3 Results of wave-amplification at 100 THz. a** Amplitude variation of the input wave through optimization time of 20 ps at the bandpass filter output; **b** Input wave intensity spectral density measured at the bandpass filter output; inset shows zoomed spectral region around 100 THz.

**Model Validation**

The accuracy of the computational model presented in this study is tested using the available experimental formula for harmonic generation by non-linear wave mixing. The experimental formula is for the generation of a new harmonic wave with angular frequency $\omega_3$ through the mixing of two monochromatic high-intensity waves with angular frequencies $\omega_1$ and $\omega_2$ in a strongly non-linear medium. The medium generates a new harmonic signal which is the sum of the two intermixed harmonics ($\omega_3 = \omega_1 + \omega_2$). The numerical efficiency for sum harmonic frequency generation is a ratio of the total wave intensity of the new harmonic $\omega_3$ to the total wave intensity of the harmonic $\omega_2$. Since both waves are of high-intensity, and our aim is not amplification, we can treat the total wave as the pump wave. We obtain numerical efficiency by solving the discretized equations for the pump wave (Eqs.1-2). The experimental formula for sum-harmonic generation efficiency is given with Eq. 10[24,25]. The variables in Eq. 10 used for the computational model validation are as follows: nonlinear coefficient d, intrinsic impedance η, speed of light in the vacuum c, the amplitude of the second wave $A_2$, the refractive index of the medium n (3.46), and length of the medium L (3.3 µm).

$$\eta_{experimental} = \frac{\omega_3}{\omega_2}\left(\sin\sqrt{2d^2\eta^3\omega_3^2(cn\varepsilon_0 A_2^2)L^2}\right)^2, \quad (10)$$

To compute and compare the numerical and the experimental efficiencies, a high-intensity wave at 180 THz is intermixed with a high-intensity wave at 120 THz. The amplitudes of the waves are $A_2$ (swept from $5\times 10^8$ V/m to $2.5\times 10^9$ V/m with an increment of $0.5\times 10^9$ V/m) and $A_1$ ($A_2/10$). The parameters of the medium are given as three resonance frequencies of the medium are 1000 THz, 1200 THz, and 1500 THz. Polarization damping rates of the medium at three resonance frequencies are 3 THz, 1 THz, and 2 THz, respectively. The relative permittivity of the interaction medium $\varepsilon_r$ is 12, and permeability of the interaction medium $\mu_r$ is 1. Resonance strengths are 1/3 for all three resonance frequencies. For validation of our formulations, we performed a simulation for a duration of 30 ps.

After calculating the numerical efficiency as $3.7 \times 10^{-4}$ for a sample pump wave amplitude of $10^9$ V/m, we estimated the nonlinearity coefficient d by equalizing the numerical efficiency to the experimental efficiency. We obtained a nonlinear coefficient as $3.31 \times 10^{-23}$ C/V$^2$. Since this estimated nonlinear coefficient is based on a single sample pump wave amplitude, we must double-check our computational model accuracy by comparing the efficiencies for some other sample pump wave amplitudes. This comparison is shown in Table 3, and we found the low error percentages. We conclude that the nonlinearity coefficient is accurately estimated. Finally, we compared the numerical and the experimental efficiencies for a much larger sample of pump wave amplitudes swept from $1 \times 10^8$ V/m to $2.5 \times 10^9$ V/m in increments of $0.5 \times 10^8$ V/m. The comparison is plotted in Fig. 4. The results are in good agreement, and the accuracy of our formulations is greater than 99%.

**Table 3 Comparing numerical and experimental sum-harmonic generation efficiencies for the nonlinearity coefficient of $3.31 \times 10^{-23}$ C/V$^2$.**

| Pump wave Amplitude (V/m) | Experimental efficiency | Numerical efficiency | Accuracy (%) |
|---|---|---|---|
| $0.5 \times 10^9$ | $9.25 \times 10^{-5}$ | $9.27 \times 10^{-5}$ | 99.8 |
| $1.0 \times 10^9$ | $3.68 \times 10^{-4}$ | $3.70 \times 10^{-4}$ | 99.5 |
| $1.5 \times 10^9$ | $8.32 \times 10^{-4}$ | $8.25 \times 10^{-4}$ | 99.2 |
| $2.0 \times 10^9$ | $1.48 \times 10^{-3}$ | $1.48 \times 10^{-3}$ | 99.7 |
| $2.5 \times 10^9$ | $2.31 \times 10^{-2}$ | $2.31 \times 10^{-2}$ | 99.5 |

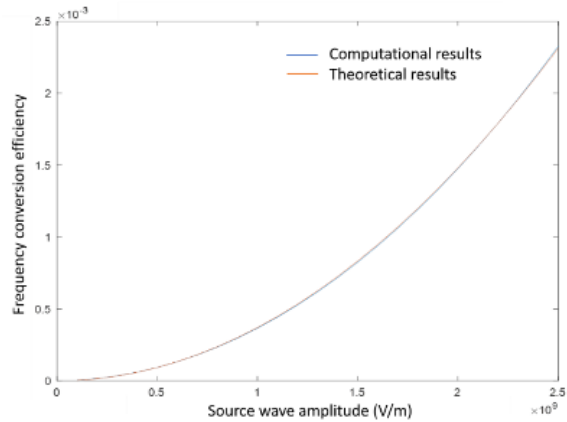

**Fig. 4 Comparison of sum-harmonic generation efficiencies for a wave at 300 THz and nonlinearity coefficient of $3.31 \times 10^{-22}$ C/V$^2$.**

**Conclusion**

Here we present novel mathematical formulations to allow energy transfer from the pump wave to the input wave. Our formulations enable us to amplify the input wave within Fabry-Perot micro-resonators. Using the Fletcher-Reeves algorithm we compute the gain of waves within the micro-resonators having multiple resonance frequencies. The results show that gains of the waves are proportional to energy density within the micro-resonators. We obtained 99% accuracy with our formulations compared to the results of well-known experimental sum harmonic generation efficiency. The mathematical formulations presented here are used to generate optical parametric amplifiers for integrated circuits, MEMS devices, and optical communications.